# To the Theory of Unstable States

Sergei G. Matinyan[*] and Mark E. Perel'man[**]

*). North Carolina Central University, 1201 Mary Townes Science Complex, Durham, NC 27707; Yerevan Physics Institute, Armenia, 375036.
.
**). Racah Institute of Physics, Hebrew University, Jerusalem, Israel.

Deviations of the decay law from exponents are discussing for a long time, however, experimental proofs of such deviations are absent. Here in the general form is shown that the conclusions about non-exponential contributions are due to the disregarding of advanced interactions, i.e. at principally non-relativistic considerations. We consider decay processes in the frame of interactions duration of the quantum field theory .We show that at this basis the usual exponential decay has place.

The time evolution of the unstable systems is usually discussed in the frame of Weisskopf-Wigner approximation [1, 2, 3], which ascribes the main properties of the decay law to a simple pole located on the second sheet of the complex energy plane (cf. [4, 5]). This yields the Fock – Krylov theorem [6] for probability of decay of quasi-stationary state as

$$P(t) = |L(t)|^2 = \left| \int \exp(-iEt/\hbar) dW(E) \right|^2, \qquad (1)$$

where $dW(E) = w(E)dE$ is the energy spectrum of the initial state. However, it does not determine the limits of integration.

The simplest cases of decay are described by the non-relativistic one-particle Breit-Wigner amplitude with the length of corresponding state vector

$$L^D \equiv \langle \Psi^D | \Psi^D \rangle = \frac{1}{\pi} \int d\omega \frac{\Gamma/2}{(\omega - \omega_0)^2 + \Gamma^2/4}. \qquad (2)$$

The integration in (2) was usually suggested as extended from $-\infty$ till $+\infty$ with $L^D(-\infty,+\infty)=1$ and correspondingly with the common exponential law of decay.

Khalfin [7] and then many others had underlined that as negative frequencies are unphysical, the integration in (2) must goes over the interval $[0, +\infty)$ (see e.g. the review [8], there are many applications of this theorem to problems of particles decay, e.g. [9]).

The integration of (2) over positive frequencies only leads to the expression:

$$L^D{}_1(0,+\infty) = \frac{1}{2} + \frac{1}{\pi} \tan^{-1}\left(\frac{2\omega_0}{\Gamma}\right), \qquad (3)$$

that goes to unity at $\Gamma \to 0$ only. This feature may be considered as a peculiar heuristic observation: the consideration based on (3) is not complete, it must be continued till completion of decay process, when become possible achieve the state that does not depend on $\Gamma$, i.e. when the processes of decoherence was ended.

Physically this expression leads to non-exponential types of decay for beginning and far times. But if a deviation from the exponential law of decay can be assumed for time close to a moment of system preparation, a deviation for very far times has not any physical justification (e.g. the system can contain at far times comparatively isolated centers only). A number of executed experiments does not fix such deviations (cf. however the recent publication [10] and its discussion [11]).

It seems that this discrepancy is due to the restriction of consideration by non-relativistic arguments without complete analysis; it must be underlined that the requirement of frequencies positivity is of classical type: QFT contains contributions of negative frequencies along with positives. In a slightly another words, it can be seen that this common approach contains retarded interactions without taking into account advanced possibilities.

Let us examine this problem via analyses of temporal properties of system.

The integrand of (2) corresponds to the delay duration at scattering (Wigner-Smith formulae, general non-relativistic theory [12]):

$$\tau_1(\omega) = \frac{\Gamma/2}{(\omega-\omega_0)^2 + \Gamma^2/4}, \qquad (4)$$

and (2) can be interpreted as the mean duration of scattering process (cf.. [13]).

But there exists the second temporal quantity, the duration of final state formation

$$\tau_2(\omega) = \frac{\omega-\omega_0}{(\omega-\omega_0)^2 + \Gamma^2/4} \qquad (5)$$

that is not taken into account in (2) and therefore any consideration of the problem without delay due to the formation of the final states can be non complete.

Therewith, we shall try to determine the course of decay by consideration of durations in general form:

$$\tau(\omega) = \frac{\partial}{i\partial\omega} \ln S(\omega), \qquad (6)$$

where $S(\omega)$ is the amplitude of elastic scattering (it means that we follow the Weisskopf-Wigner approach). As $S(\omega) = |S(\omega)| \exp\varphi(\omega)$, the expression for durations can be represented as

$$\tau(\omega) \equiv \tau_1 + i\tau_2 = \frac{\partial}{\partial\omega}\varphi(\omega) + \frac{\partial}{i\partial\omega}\ln|S(\omega)|. \qquad (6')$$

The general definition (6) leads to (4) and (5) for the case of single pole. Its Fourier transformation,

$$\tau(t) = \frac{1}{2\pi}\int_{-\infty}^{\infty} d\omega e^{i\omega t} \tau(\omega) = \frac{1}{2\pi i}\int_{-\infty}^{\infty} d\omega e^{i\omega t} \frac{\partial}{\partial\omega}\ln S(\omega). \qquad (7)$$

It must express temporal evolution of decay via the Fourier transform of logarithmic residue of S-matrix with physically substantiated limits of integration.

Corresponding integrals are of such general form (e.g. [14]):

$$\frac{1}{2\pi i}\int_C dz\,\varphi(z)\frac{d}{dz}\ln\phi(z) = \frac{1}{2}\sum\left(n_k\varphi(a_k) - p_k\varphi(b_k)\right), \qquad (8)$$

where C is the suitable closed contour, $a_k$ and $b_k$ are residues in zeros and poles, $n_k$ and $p_k$ are their repetition numbers.

The analyticity of the causal $S(\omega)$ and the condition of unitarity $S(-\omega) = S^*(\omega)$ allow its sufficiently general representation as the Bläschke product:

$$S(\omega) = const \cdot |\omega|^{-p} \prod_n \frac{\omega - \omega_n + i\gamma_n/2}{\omega - \omega_n - i\gamma_n/2} \times \frac{\omega + \omega_n - i\gamma_n/2}{\omega + \omega_n + i\gamma_n/2} \Rightarrow S^{(=)}(\omega) \times S^{(+)}(\omega). \qquad (9)$$

Second multipliers in (9) can be evidently omitted at calculation of non-relativistic resonance reactions with positive $\omega$ since corresponding terms give too small contributions.

But at the consideration of the general problems it must be taken into account that the causal (Feynman) propagator includes positive and negative frequencies functions on the equal basis: e.g. $\Delta_c(x) = \theta(t)\Delta^{(-)} - \theta(-t)\Delta^{(+)}$ or similar representation with account of $\Delta_{R,A}$.

From the unitarity of $S(\omega)$ follows that $\tau(-\omega) = \tau^*(\omega)$, i.e. the positions of zeros and poles on the complex energy plane are not varied. It means that the first and the second quadrants of the complex energy plane give similar contributions into (7) that can be evaluated by simple closing of the counter by big half circle. By such a way the condition ($0 \leq \omega < \infty$), very awkward for quantum theory, must be replaced by the more simple and common condition ($-\infty < \omega < \infty$) for integrals (7).

These integrals evidently lead after averaging to the usual expression:

$$\bar\tau(t) \longrightarrow \sum_n \exp(-\gamma_n t) \qquad (10)$$

Notice that the experimentally established dependence of $t^2$ of decay rate at initial moments [15, 16] can be attributed to a reversible process, which contradicts an irreversible decay to the continuum and must be considered separately.

The analyticity of $S(\omega+i\varsigma)$ in the upper half-plane allows to write instead of (7) such integral over the closed contour:

$$\oint \tau(\omega)d\omega = \oint \tau_1(\omega)d\omega = 2\pi(N-P), \qquad (11)$$

where N and P are zeros and poles of temporal function inside the contour. Poles of $\tau_1(\omega)$ signify impossibility of signal transferring on these frequencies through the system (frequencies locking) or particles capture at scattering processes. Zeros show that corresponding signals are passed through system without delays, etc. Really (11) represents a variant of the Levinson theorem of quantum scattering theory, e.g. [17].

The maximum-modulus principle for $S(\omega)$ shows that as $\tau_2(\omega)$ is determined via its derivative, it can not be equal to zero at any frequency: the formation of outgoing signal (wave, particle, state) always requires some temporal duration.

Let us note that the phase transition of the first kind in more ordered states, at constant temperature and pressure, at least, can be considered as the decay process with emission of latent heat [18] and is also describable via temporal functions [19].

Comparison of resonance scattering and decays amplitudes initially had difficulties, since processes with different number of particles in in- and out-states requires the rigged Hilbert spaces [20] (cf. [21]).

In conclusion we can underline that the above considerations evidently show the usefulness and significance of the temporal approach. It also shows that the small relativistic contributions can be present at obviously non-relativistic problem.

------------------------------------

**REFERENCES**

E-mails:  smatinian@nc.rr.com; m.e.perelman@gmail.com.

[1] . G. Gamow, Z. Phys. **51** (1928) 204.

[2]. V. Weisskopf and E.P. Wigner, Z. Phys. **63** (1930) 54.

[3]. G. Breit and E.P. Wigner, Phys. Rev. **49** (1936) 519.

[4] . J. Schwinger, Annals of Physics **9** (1960) 169.

[5]. M. Veltman, Physica **29** (1963) 186.

[6] . V. Fock and N. Krylov, J. Phys. **11** (1947) 112.

[7] .  L.A. Khalfin, Dokl. Acad. Nauk USSR **115** (1957) 277 [Sov. Phys. Dokl. **2** (1957) 340].

[8] . L. Fonda, G. C. Ghirardi and A. Rimini, Rep. Prog. Phys. **41** (1978) 587.

[9] . K. Urbanowski. In arXiv:/0712.0328; /0803.31.88 and references therein.

[10] . C. Rothe, S. I. Hintschich, and A. P. Monkman, Phys. Rev. Lett. **96** (2006) 163601.

[11] . J. Martorell, G. J. Muga and D. W. L. Sprung, In: arXiv /0709. 2685.

[12] .  M. E. Perel'man, Int. J. Theor. Phys., **47** (2008) 468.

[13] . L. Maiani and M. Testa,  In: arXiv/9709110.

[14] .  M. A. Lavrent'ev amd B. V. Shabad, Metodi Teorii Fynkzij Kompleksnogo Peremennogo, 4th edn., Moscow: Nauka, 1973.

[15] . V.J. Menon and A.V. Lagu, Phys. Rev. Lett., **51** (1983) 1407.

[16]. R. Wilkinson et al.  Nature, **387** (1997) 575.

[17] . M.L. Goldberger and K.M. Watson, Collision Theory. Wiley: New York (1964).

[18] . M.E. Perel'man and V.A. Tatartchenko, Phys. Lett. A **372** (2008) 2480.

[19] . M.E. Perel'man, Phil. Mag., **87** (2007) 3129.

[20] . M.E. Perel'man, Bull. Acad. Sc. Georgian SSR  **62** (1971) 33 (Main part in English: Math. Rev. **47** (1974) 1776).

[21] . A  Bohm, P. Bryant and Y. Sato. In: arXiv/0803-32333.